\documentclass[traditabstract]{aa}

\usepackage{graphicx}
\usepackage{txfonts}
\usepackage{hyperref}

\usepackage{amsmath}
\usepackage{amssymb}
\usepackage{graphicx}
\usepackage{newtxtext,newtxmath}
\usepackage{Times}

\usepackage[T1]{fontenc}
\usepackage{ae,aecompl}

\usepackage{booktabs}
\usepackage{lipsum}
\usepackage{multirow}
\usepackage{natbib}
\usepackage{soul}
\usepackage{textcomp}

\newcommand{\citeprep}[1]{{\color{blue}#1 (in prep.)}}
\newcommand{\citeprepp}[1]{{(\color{blue}#1 in prep.)}}

\newcommand{\msun}[1]{$10^{#1}\mathrm{M_\odot}$}
\newcommand{\zpeg}{\textsc{zpeg}}
\newcommand{\lephare}{\textsc{LePhare}}
\newcommand{\cigale}{\textsc{cigale}}
\newcommand{\magphys}{\textsc{magphys}}
\newcommand{\prospector}{\textsc{prospector}}

\makeatletter
\newcommand*{\citelink}[1]{\hyper@link{cite}{cite.#1}}

\makeatother

\bibpunct[ ]{(}{)}{;}{a}{}{,}

\hypersetup{
     colorlinks   = True,
     citecolor    = blue,
     linkcolor = blue,
     urlcolor = blue
}

\newcommand\deltaOm{0.001}
\newcommand\deltaOmpercent{0.3}
\newcommand\deltaM{0.004}
\newcommand\deltaMpercent{6}
\newcommand\deltaw{0.006}
\newcommand\deltawpercent{0.6}
\newcommand\deltaH{0.2}
\newcommand\deltaHpercent{0.3}

\begin{document}


\title{Systematic errors on optical-SED stellar mass estimates for galaxies across cosmic time and their impact on cosmology}

\author{
    A. Paulino-Afonso \inst{1}\fnmsep\thanks{E-mail: apaulinoafonso@tecnico.ulisboa.pt} \and
    S. Gonz\'alez-Gait\'an\inst{1} \and
    L. Galbany \inst{2,3} \and
    A. M. Mourão \inst{1} \and
    C. R. Angus \inst{4} \and
    M. Smith \inst{5} \and
    J. P. Anderson \inst{6} \and
    J. D. Lyman \inst{7} \and
    H. Kuncarayakti \inst{8,9} \and
    M. A. Rodrigues \inst{10}
}

\institute{CENTRA, Instituto Superior T\'ecnico, Universidade de Lisboa, Av.  Rovisco Pais 1, 1049-001 Lisboa, Portugal
    \and
    Institute of Space Sciences (ICE, CSIC), Campus UAB, Carrer de Can Magrans, s/n, E-08193 Barcelona, Spain.
    \and
    Institut d'Estudis Espacials de Catalunya (IEEC), E-08034 Barcelona, Spain.
    \and
    DARK, Niels Bohr Institute, University of Copenhagen, Jagtvej 128, 2200 Copenhagen, Denmark
    \and
    School of Physics and Astronomy, University of Southampton, Southampton, SO17 1BJ, UK
    \and 
    European Southern Observatory, Alonso de C\'ordova 3107, Casilla 19, Santiago, Chile
    \and 
    Department of Physics, University of Warwick, Coventry, CV4 7AL, UK  
    \and 
    Tuorla Observatory, Department of Physics and Astronomy, FI-20014 University of Turku, Finland
    \and
    Finnish Centre for Astronomy with ESO (FINCA), FI-20014 University of Turku, Finland
    \and
    GEPI, Observatoire de Paris, PSL University, CNRS, 5 Place Jules Janssen, F-92190 Meudon, France
}

\date{ }

\abstract{
    Studying galaxies at different cosmic epochs entails several observational effects that need to be taken into account to compare populations across a large time span in a consistent manner. We use a sample of 166 nearby galaxies that hosted type Ia supernovae (SNe Ia) and have been observed with the integral field spectrograph MUSE through the AMUSING survey.  Here, we present a study of the systematic errors and bias in the host stellar mass with increasing redshifts that are generally overlooked in SNe Ia cosmological analyses.  We simulate observations at different redshifts ($0.1 < z < 2.0$) using four photometric bands ($griz$, similar to the Dark Energy Survey-SN program) to then estimate the host galaxy properties across cosmic time. We find that stellar masses are systematically underestimated as we move towards higher redshifts, due mostly to different rest-frame wavelength coverage, with differences reaching 0.3 dex at $z\sim1$. We have used the newly derived corrections as a function of redshift to correct the stellar masses of a known sample of SN Ia hosts and derive cosmological parameters. We show that these corrections have a small impact on the derived cosmological parameters.  The most affected is the value of the mass step $\Delta_M$, which is reduced by $\sim$\deltaM\ (\deltaMpercent\% lower).  The dark energy equation of state parameter $w$ changes by $\Delta w \sim $ \deltaw\ (\deltawpercent\% higher) and the value of $\Omega_m$ increases at most by \deltaOm\ ($\sim$\deltaOmpercent\%),  all within the derived uncertainties of the model. While the systematic error found in the estimate of the host stellar mass does not significantly affect the derived cosmological parameters, it is an important source of a systematic error that one should correct for as we enter a new era of precision cosmology.
}


\authorrunning{A. Paulino-Afonso et al.}
\titlerunning{Galaxy stellar masses estimates and its impact on cosmology}

\maketitle

\section{Introduction}\label{section:introduction}

Type Ia supernovae (SNe Ia) have been successful as standard candles to probe the expansion history of our Universe over the last decades \citep[see e.g.][]{riess1998, perlmutter1999, betoule2014, riess2018, scolnic2018, descollaboration2019}. However, SNe Ia are not perfect standard candles, and several empirical corrections are used to estimate their intrinsic luminosity. For example, light-curve shapes \citep{phillips1993} and colours \citep{riess1996, tripp1998} have been {used} to reduce the scatter of their peak magnitudes by 50\% and improve distance errors {down to} $\sim7$\%. With increasing samples of spectroscopically confirmed \citep[e.g.][]{scolnic2018, smith2020} and photometrically classified SNe Ia \citep{jones2018a} we are now in a phase where understanding the origin of these empirical corrections will improve our constraints and provide for better corrections. This has potential implications for the determination of the equation of state of the Universe.

The observed scatter of SNe Ia distance residuals for the best-fit cosmological model is close to the 0.1 mag level \citep[see e.g.][]{brout2019b}. This indicates that either there is a limit to which one can standardize SNe Ia, or there are additional correlations to their peak brightness that are not yet known due to limits on the quality of existing samples. These additional correlations are thought to arise from uncertainties related to the {progenitor properties,} physics of SNe Ia explosions and/or the environment in which they occur \citep[see, e.g.][]{scannapieco2005, mannucci2006, maoz2014, livio2018}. The drive to obtain ever more accurate standardizations of SNe Ia has motivated the search for additional empirical corrections based on the properties of the host galaxy used as a tracer of the SNe Ia progenitors \citep[e.g.][]{hicken2009, sullivan2010, kelly2010,lampeitl2010,gupta2011, dandrea2011, hayden2013, rigault2013, childress2013, johansson2013,pan2014, uddin2017, uddin2020, ponder2020,smith2020}.

One of the most commonly used empirical corrections is based on the host stellar mass, with studies finding that SNe Ia occurring in galaxies with $M_\star>10^{10}\mathrm{M_\odot}$ {require additional brightness corrections} compared to those found in lower stellar mass galaxies \citep[e.g.][]{sullivan2010, kelly2010,lampeitl2010}. Such a correction has been found in multiple studies, at various degrees of confidence ($3 - 6 \sigma$) using multiple samples in the low and high-redshift Universe \citep[e.g.][]{sullivan2010, kelly2010,lampeitl2010, childress2013, johansson2013,pan2014, uddin2017,uddin2020,ponder2020}.  However,  it has been shown that more recent fitting frameworks lead to reduced corrections \citep[e.g.][]{brout2019b, smith2020}. There is currently no consensus on the physical motivation for this correction, as the stellar mass of galaxies is found to correlate with other global properties of the host galaxy: star-formation rate \citep[e.g.][]{speagle2014}, metallicity \citep[e.g.][]{tremonti2004, curti2020}, and dust \citep[e.g.][]{garn2010}. Thus, it has also been found that the excess scatter could be corrected using other physical parameters of the host galaxy such as their metallicity and stellar age \citep{gupta2011, dandrea2011, hayden2013,pan2014,moreno-raya2016}, star-formation rate \citep{sullivan2010} or dust \citep[][]{brout2021}.

The studies mentioned above focused on the global properties of the host galaxy since, for large cosmological distances, these are the only possible measurements with current instrumentation. Nonetheless, the progenitors of SNe Ia might reside in a particular region of the galaxy that is not well traced by their global properties. Recent studies on {nearby} galaxies have traced the empirical corrections to the local environment in which the SNe Ia occur {\cite[][]{stanishev2012, rigault2013,rigault2015,rigault2020, galbany2014, galbany2016b, jones2015,jones2018b, moreno-raya2016,roman2018,kim2018, kim2019,rose2019,rose2021a,kelsey2021}.} In these studies, the authors focused on the local star formation rate (traced by H$\alpha$ emission or local $U-V$/$u-g$ colours) to find that SNe Ia in actively star-forming environments are fainter than those found in more passive environments.  However,  \cite{jones2015} and \cite{jones2018b} find no conclusive evidence that {correlations built from the local properties} are better than those found with global properties.

Despite the existence of different empirical corrections, that based on the global host stellar mass has been the mostly used in cosmological analysis using SNe Ia \citep[e.g.][]{sullivan2011, betoule2014, scolnic2018,popovic2021}. This is a consequence of the stellar mass being a more straightforward measurement to obtain, as it is the most robust parameter that can be estimated from photometry alone \citep[e.g.][]{pforr2012}. Nonetheless,  care should be taken when estimating stellar masses and comparing estimates across a large redshift range,  especially when using  a small number of photometric bands as is typical in photometric studies of SNe. In this scenario one needs to account for observational effects (cosmological dimming, rest-frame coverage) that can impact the derived parameters.  We aim to quantify the systematic errors on the estimates of stellar masses from the same photometric bands across a large redshift range,  and test its impact on the derived cosmological parameters from supernovae studies.

In this paper, we use a sample of 166 nearby galaxies with integral field spectroscopic (IFS) data from the \emph{All-weather MUse Supernova Integral field Nearby Galaxies} (AMUSING) survey \citep[][]{galbany2016} to simulate photometric observations of {the same galaxies between} $0.1 < z < 2.0$.  {Using our host galaxy IFS data, we have simulated \emph{griz} observations and derived the host galaxy properties with commonly used} spectral energy distribution (SED) fitting codes.  {We then take} the observed differences between the new simulated properties and those derived in the local Universe to estimate a redshift-dependent stellar-mass correction. We use our new correction in our cosmological analysis and show the impact on the derived cosmological parameters.

This manuscript is organized as follows: in Section \ref{sec:data} we briefly explain the AMUSING survey on which our manuscript is based.  {In Section \ref{sec:artredshift} we explain our novel method for simulating galaxy observations at higher redshift.  Section \ref{sec:measure_mass} details the different stellar mass estimates that are used throughout the paper.} We show our results regarding systematic errors on stellar mass estimation and their impact on the derivation of cosmological parameters, and we discuss our findings within the current $\Lambda$CDM paradigm in Section \ref{sec:results}. We summarize our main conclusions in Section \ref{sec:conclusions}. We use AB magnitudes \citep{oke1983}, a Chabrier \citep{chabrier2003} initial mass function (IMF) unless otherwise explicitly stated, and assume a $\Lambda$CDM cosmology with H$_{0}$=70 km s$^{-1}$Mpc$^{-1}$, $\Omega_{M}$=0.3, and $\Omega_{\Lambda}$=0.7.


\section{The AMUSING survey}\label{sec:data}

\begin{figure}
    \centering
    \includegraphics[width=\linewidth]{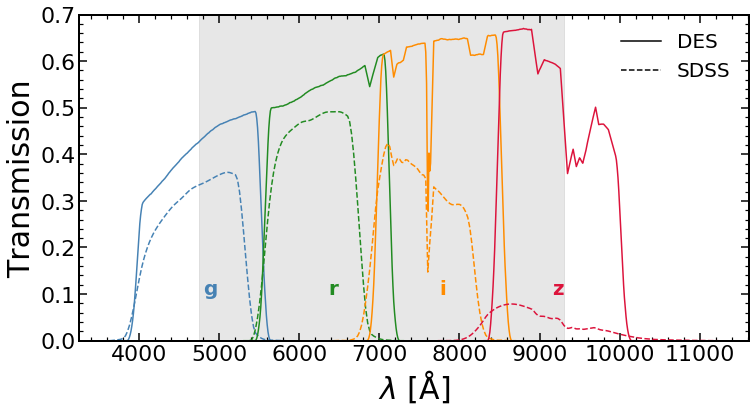}
    \caption{We highlight {with the shaded region} the coverage of the MUSE spectroscopic data in comparison to the coverage of the DECam and SDSS \emph{griz} filters.}
    \label{fig:filters}
\end{figure}

In this work we use a sample of SN host galaxies drawn from the AMUSING survey \footnotemark\footnotetext{Based on observations made with ESO Telescopes at the Paranal Observatory (programmes 95.D-0091(A/B), 96.D-0296(A), 97.D-0408(A), 98.D-0115(A), P99 - 099.D-0022(A), P100 - 100.D-0341(A), P101 - 101.D-0748(A/B), P102 - 102.D-0095(A), P103 - 103.D-0440(A/B) and P104 - 104.D-0503(A/B).} \citeprep{Galbany et al.}. Data were obtained with the Wide Field Mode of the MUSE instrument \citep{bacon2010} installed at the UT4 of the Very Large Telescope in Chile. Each pointing has approximately 1\arcmin × 1\arcmin field of view (FoV) taken at a scale of 0.2\arcsec/pixel. The spectra have a wavelength coverage in the optical range (4750\AA-9300\AA, see Fig. \ref{fig:filters} for a comparison with the DECam and SDSS $griz$ filter sets) with a fixed spectral sampling of 1.25\AA\ (spectral resolution of around 1800 at the blue edge and 3600 at the red edge).  Our observations have a median seeing of $\sim$1\arcsec\, which corresponds to a physical resolution around $\sim600$pc at the median redshift of our sample, $<z>=0.03$ {(with 75\% of the sample below $z=0.05$), corresponding to a distance of $\sim$124 Mpc}.

The data used in our work has been reduced using the MUSE pipeline \citep[v1.2.1, ][]{weilbacher2014} and the Reflex environment \citep{freudling2013}.  Tasks performed by the pipeline include standard reduction  such as subtracting bias, flat fielding, galactic extinction corrections, and flux/wavelength calibrations.  For removal of the sky background,  we use either an offset pointing to an empty region or blank sky regions within the science frames themselves (for smaller targets) and use the Zurich Atmosphere Purge package \citep[ZAP,][]{soto2016} to perform this task.  To reconstruct the final data product we applied a geometrical transformation of the individual slices to align them in a datacube.  For more information on this procedure we refer to \citet{galbany2016} and \citet{kruhler2017}. We have further corrected the fluxes of the observed spectra by matching the flux of the integrated galaxy light in the $r$-band to, by order of priority of available data, Pan-STARRS, DES, and SDSS photometry  \citeprepp{Galbany et al.}.

\subsection{Our sample}

Our study is based on a subsample of the AMUSING survey that {selects} only {SNe Ia host} galaxies for which the FoV covers the entire galaxy,  and no significant {foreground contamination by bright stars or background contamination by distant galaxies} is found in the MUSE datacubes. No galaxies with $z\geq0.1$ are selected, with the great majority ($\sim75$\%) having $z<0.05$. There is no additional cut on any other property within the sample. And since the existence of foreground stars and/or background galaxies does not depend on either the host galaxy or the SNIa,  the resulting subsample is akin to a random sampling of the parent sample. This process was conducted through visual inspection of each object and its corresponding segmentation map. This map is defined as {the selection of} all pixels belonging to the object of interest flagged, and it was done as a combination of two steps.

First, we searched for Gaia matches within the field-of-view of the MUSE datacube with a 1\arcmin\ radial search around the cube centre using the \texttt{astroquery} package \citep{astroquery}. Then we selected as foreground stars all objects with good parallax ($\pi$) measurements (i.e. $\pi > 2\pi_\mathrm{err}$ {with $\pi_\mathrm{err}$ being the error on the parallax}). With the final list of foreground stars, we built individual circular masks centred on each and with a radius containing 95\% of the flux measured within a 3\arcsec radius. Then, we select as the final object map all connected spaxels with {a S/N} > 3 that belong to the target object and do not overlap with the circular masks defined in the previous step. {A similar S/N cut is applied when measuring photometry in the simulated observations (see Section \ref{sec:artredshift}).}

\begin{figure}
    \centering
    \includegraphics[width=\linewidth]{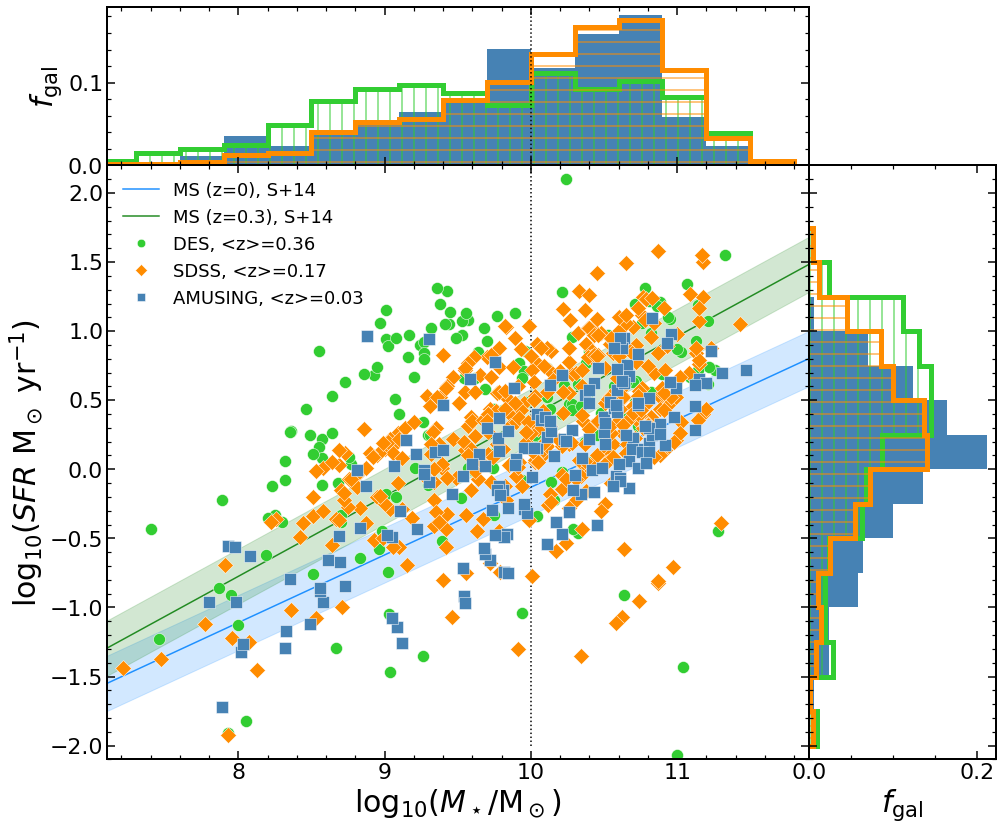}
    \caption{The comparison of the AMUSING sample stellar masses and star-formation rates computed using \textsc{magphys}, see Section \ref{sec:measure_mass} (in blue), with the sample from \cite{smith2020} (in green) and the one from \cite{sako2018} (in orange). We show as lines with shaded regions the expected relation between stellar mass and SFR (commonly referred to as 'Main-Sequence') for the population of star-forming galaxies at different redshifts (adapted from \citealt{speagle2014}).  {We show in the upper panel the stellar mass distributions and in the right panel the $SFR$ distributions for the three samples.} We highlight the $10^{10}\mathrm{M_\odot}$ threshold for SNe Ia brightness corrections {\citep[see e.g.][]{sullivan2010}} as the vertical dotted line.}
    \label{fig:mass_sfr}
\end{figure}

All segmentation maps were individually inspected to select only objects without clear interlopers and with no other nearby objects {(either bright foreground stars or background galaxies)} that may contaminate the light of the galaxy of interest.  After this inspection, a total of 166 galaxies were selected to be included in our study.  {To establish a comparison with other host galaxy samples in the literature, we have computed the physical properties (stellar masses and star-formation rates) of our AMUSING subsample using \textsc{magphys}, as described in section \ref{sec:measure_mass}, and a \textit{griz} magnitude set (using any of the other codes described below does not change significantly the results.  This is comparable to the stellar mass estimates of the  SDSS \citep[][{\!, $<z>=$0.17}]{sako2018} and DES-SN program \citep[][{\!, $<z>=$0.36}]{smith2020} samples.} \footnotemark\footnotetext{{We computed stellar masses using both their published catalog photometry and \textsc{magphys} and find negligible differences to their published values (smaller than 0.05 dex).}} As we show in Fig. \ref{fig:mass_sfr},  the AMUSING sub-sample spans similar stellar mass ranges as the samples from SDSS and has more massive galaxies on average than the sample from DES-SN program. {This latter difference could be naturally explained by different cosmic epochs probed by the two samples. Our AMUSING lower redshift sample galaxies would have had more time to build up their stellar masses.}  In terms of star-formation rates, we find that our sample is slightly less star-forming on average than the other two programs, but that can be easily explained by the median redshift of the sample, as one expects galaxies to increase their star-formation as we move from $z\sim0$ to $z\sim2$ \citep[see e.g.][]{madau2014, speagle2014}.  {This shows that our population of galaxies is not particularly biased, and the differences among different surveys can be attributed to the different redshift ranges that are being targeted.}

\section{Reconstructing data cubes at higher redshift}\label{sec:artredshift}

\subsection{Extending the MUSE datacubes}

{We are interested in testing the impact of the observed wavelength range on estimated stellar masses. To simulate observations in a large redshift range we need to have an extended wavelength baseline. We note, however, that for galaxies in our AMUSING subsample the differences in rest-frame coverage form one galaxy to another are negligible ($\Delta z \approx 0.03$) and much smaller than what we aim to simulate in our work.}

\begin{figure*}
    \centering
    \includegraphics[width=\linewidth]{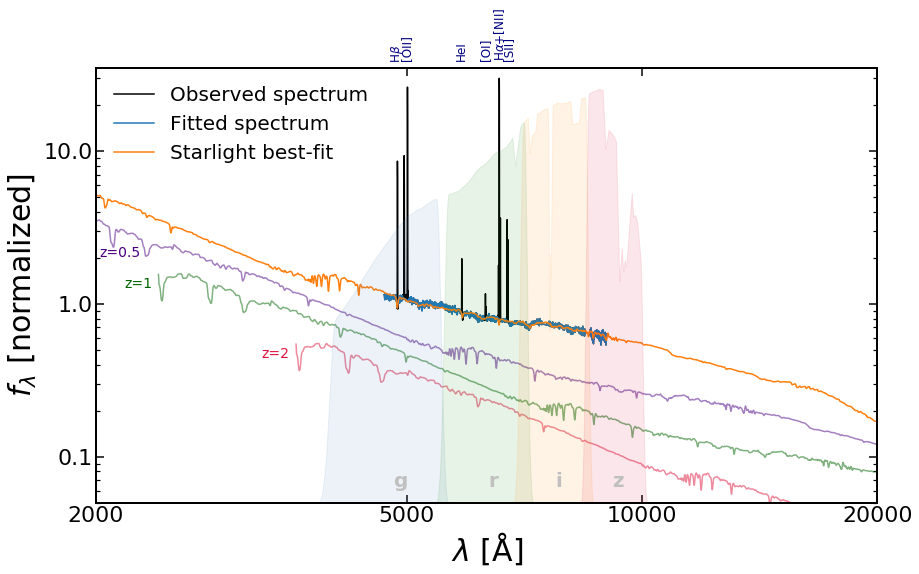}
    \caption{Example of {a spectrum with an} extended wavelength coverage {obtained from} the best-fit \textsc{starlight} model (in orange) compared to the {original} MUSE data (in black). The best-fit is done on the masked spectra (in blue).  {The transmission curves of the DES filters are shown as shaded regions.} We also show the observed wavelengths of the redshifted spectra at z=0.5, 1, and 2 in this figure as the purple, green, and red lines, respectively. Vertical offsets were applied for better visualization. On top of the plot we identify the observed strong emission lines.}
    \label{fig:starlight_fit}
\end{figure*}

To perform a simulation of the galaxy spectral energy distribution (SED) using a broad range of filters, we thus artificially extended the data available in the MUSE datacubes (which covers the region 4750-9000\AA) to span a larger {rest-frame} wavelength coverage: 1200\AA\ - 20000\AA. To do so we use \textsc{starlight} \citep[][]{fernandes2005} to perform a spaxel-by-spaxel fit of the local spectra and then use the best-fit model to get the extended wavelength coverage (see Fig. \ref{fig:starlight_fit}).  {Prior to the fit with \textsc{starlight}, all major emission lines are masked {as none of the models include them} (the blue line in Fig. \ref{fig:starlight_fit})}. {We expect that the masking of emission lines will have negligible impact on the derived stellar masses, which is the main goal in our work \citep[e.g][]{whitaker2014}.} This fit is done for all spaxels belonging to each object map as defined in the previous section. The choice of the extended coverage {takes into account that} simulated galaxies {will be} used with optical and near-infrared filters across a large redshift range ($z\lesssim2$).

We use a combination of 45 base spectra built with the \citet{bruzual2003} library and a \citet{chabrier2003} initial mass function. The base spectra span 15 {stellar} ages from 1 Myr to 13 Gyr and three metallicities (Z = 0.004, 0.02 and 0.05). {The best-fit SSP template is then constructed as a linear combination of these base spectra that best approximates the observed spectra}.

\subsection{Artificial redshifting of galaxies}

To estimate how the perceived properties of galaxies change across cosmic time, we wrote an algorithm (hereafter referred to as \textsc{argas}) to simulate observations of how galaxies in the local Universe would look if they were at {higher redshift}. This is done by artificially redshifting galaxies following closely the method described in \citet[][\!\!, see also \citealt{barden2008}]{paulino-afonso2017}.

The core of the algorithm consists of three separate transformations:
\begin{enumerate}
    \item We apply a flux correction {to the datacube} (the dimming factor) that scales as the inverse of the luminosity distance to the galaxy.
    \item We re-scale each wavelength slice of the cube (i.e. a 2D image at that wavelength) to match the pixel scale of the high-redshift observations whilst preserving the physical scale and flux of the galaxy.
    \item We redshift the {extended galaxy spectra of each spaxel} to match the observed frame at the requested redshift.
\end{enumerate}

We show in Fig. \ref{fig:dimm_all} the effects of the scaling and dimming on images of a galaxy for the four different filters. The same method was applied to all slices of the extended datacube to re-create a MUSE observation at higher redshifts. From this extended and redshifted datacube one can then extract photometry from filters within the {{observed} wavelength interval between  $1200\times(1+z)$\AA \, and $20000\times(1+z)$\AA} for assessing possible biases in the estimation of physical galaxy parameters from photometric data (e.g. stellar mass or star-formation rates).

We have applied each of these effects (dimming, scaling, and redshifting) separately and find that cosmological dimming is counteracted by the reduced physical resolution of higher redshift images.  {This experiment nicely confirms the concept of surface brightness which is independent of distance {for instruments with the same resolution. This occurs since, while the flux observed at higher redshift is lower due to the cosmological dimming effect, each pixel also covers a larger physical area of the galaxy which naturally corresponds to higher emitted flux per pixel. And since both the luminosity distance and angular diameter distance scale similarly with redshift, they tend to cancel each other.}  We find that we lose some flux in the  outskirts of galaxies as we move towards higher redshifts. Nonetheless, the different rest-wavelength coverage of the photometric filters has the most significant impact on the derived physical parameters. {The rest-frame coverage changes with redshift, towards bluer wavelengths as we move to higher redshifts when using the same filter set, leading to the major contribution to the observed differences.} The use of 3D data from MUSE allows for a more accurate depiction of observational effects than simply simulating integrated SEDs, as it allows to measure the impact of flux loss in galaxy outskirts due to surface brightness dimming, as well as a good handle on the observed wavelength dependence of the flux.

\begin{figure*}
    \centering
    \includegraphics[width=\textwidth]{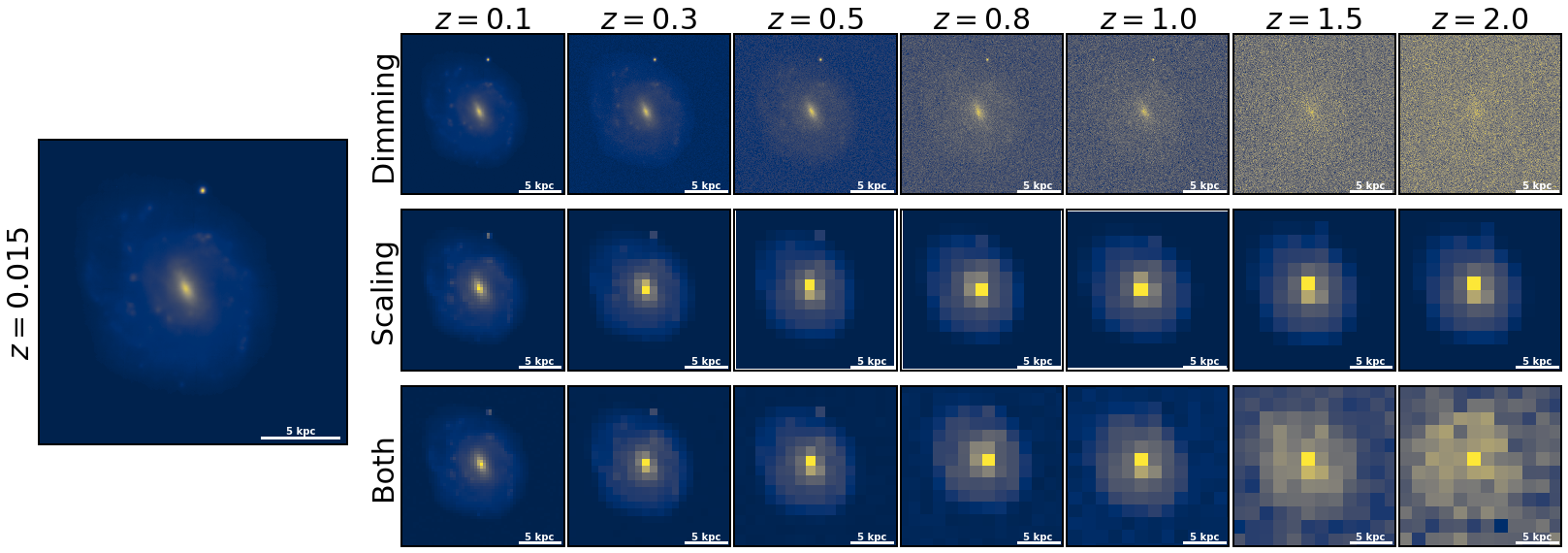}
    \caption{Example of the different observational effects {as simulated for} a single scale instrument (same as used in our simulations, with a pixel scale of 0.2\arcsec) {showing} a galaxy (PGC 128348, host of SN Ia ASASSN-14jg) simulated at various redshifts {(no SED redshifting is applied here)}. The original image is shown in the single panel on the left.  The simulated images are shown in the grid with redshifts increasing from left to right and, from top to bottom, the effects of dimming, scaling and both applied to the galaxy.  Each square has the exact same physical scale of $\approx$ 20$\times$20 kpc.}
    \label{fig:dimm_all}
\end{figure*}

\subsection{Noise addition}\label{ssec:noise}

To simulate realistic observation conditions, we need to add noise to the simulated high-redshift images. We assume that the noise is well described by a Gaussian distribution with a width defined by $\sigma_\mathrm{rms}$.  {We have tested two approaches to simulated noise.}

One approach is to scale the noise of the original MUSE datacube to the desired exposure time of the simulated observations. In doing this, we assume that the RMS is inversely proportional to the exposure time. In practice, we build a 2D noise map matching each of the filters we want to test. For each exposure time we have that for an exposure time $t$ the noise is described by a $\sigma_\mathrm{rms} (t) = \sigma_\mathrm{rms,0} \times t_0/t$, with $t_0$ and $\sigma_\mathrm{rms,0}$ being the exposure time and noise properties of the original datacube.

A second approach is to define a magnitude limit for each set of observed filters. To do this, we simulate a point-like object as a 2D Gaussian profile with an FWHM = 3 pixels {(which is the typical sampling of a PSF,  depending only on the instrument)} with a flat spectrum with a constant value $f_\star$. We determine $f_\star$ to be the value for which the integrated magnitude in the observed filter and within a 3\arcsec\ aperture is equal to the desired magnitude limit.  Then we compute the $\sigma_\mathrm{rms}$ that allows the simulated star to be detected with an S/N = 5 in the 3\arcsec\ aperture.  {This helps simulate the conditions of typical surveys, for which the limiting depth is similar across the observed fields. The value of $\sigma_\mathrm{rms}$}  is estimated by exploring a fixed list of values,  computing the magnitude of the star at each value of $\sigma_\mathrm{rms}$ and comparing that to the real magnitude of the star. Once the difference between magnitudes exceeds 0.2 mag\footnotemark\footnotetext{A S/N=5 means a 20\% error on the flux, which translates to $-2.5\log_{10}(1.2) \approx 0.2$mag.}, we select that value of $\sigma_\mathrm{rms}$ {to fix our simulated survey depth}. To remove the bias of having a particular realization of a 2D Gaussian noise distribution to define our final value of $\sigma_\mathrm{rms}$, we repeated this procedure 200 times and defined as our final value of $\sigma_\mathrm{rms}$ the median of those 200 realizations.

In the remainder of the paper, we use simulations with noise added as described in the second approach. Our choice was made since this approach is the one that can most easily be matched to existing survey designs given the publicly available information. The conclusions from our work do not change if we choose the first approach to add noise to the images. {We have simulated galaxies with 4 different limiting magnitudes, $m_\mathrm{lim}$: 25, 27, 29, and 31. The results in this work are all based on a value of $m_\mathrm{lim} = 27$ \citep[akin to the wide COSMOS survey, ][]{scoville2007,koekemoer2007} used for all redshifts. The conclusions from this work remain similar if we use any of the other three values, with the exception that we fail to detect most of the sources at $z>1.5$ when simulating with $m_\mathrm{lim} = 25$. This implies that to observe galaxies at $z<1$ using an instrument with the simulated plate scale of 0.2\arcsec/pix, it suffices to have a depth of $\sim$25 mag across all photometric bands. For $m_\mathrm{lim} = 27$, we detect $\gtrsim$90\% of the sample in all photometric bands at all redshifts.}

\section{Estimating stellar masses}\label{sec:measure_mass}

Estimating {a galaxy stellar mass} from photometric data has been a powerful driver of extragalactic studies over the past decades. In particular, SED fitting codes have often been used {with this goal} \citep[][]{borgne2002,burgarella2005,ilbert2006,cunha2008,kriek2009,carnall2018,johnson2021}. However, getting the right stellar mass estimate is not yet a well-posed problem due to the large number of model choices that one can {make} prior to fitting data \citep[e.g.][]{pforr2012, mitchell2013, acquaviva2015, mobasher2015,lower2020}. To estimate the stellar masses and {star-formation rates} ($SFR$) for the galaxies in our sample, we have performed our SED fitting using several publicly available SED fitting codes that we describe below.  {We have tried, whenever possible to use the same set of templates and configurations among different SED fitting codes, although that is not always possible due to individual code design choices.  We detail below the set of templates/choices used with each code.  All our fitting was done using the photometric data derived for DES $griz$ filter set, as seen in Fig. \ref{fig:filters}. In fitting of the SEDs, the redshift of the galaxies is known from the spectra and fixed.}

\subsection{ZPEG}\label{ssec:zpeg}

In \textsc{ZPEG} \citep{borgne2002} the template library {for the stellar populations} is built from \textsc{PEGASE.2} \citep[][]{fioc1997} from a set of nine exponentially declining star-formation histories, where
\begin{equation}
    SFR(t) \propto \frac{\exp\left(-t/\tau\right)}{\tau}
\end{equation}
with $t$ being the age of the galaxy and $\tau$, the e-folding time, a parameter with the possible values of : 0.1, 0.2, 0.3, 0.4, 0.5,0. 75, 1, 1.5,  or 2 Gyr. The SED is computed for 201 {timesteps} from 0 to 14 Gyr and the standard nebular emission prescription is used {\citep[see][for details on this]{fioc1997}.} For each template, the initial metallicity has a value of 0.004 and evolves with time (with new stars having the metallicity of the ISM).  We use the \citet{kroupa2001} IMF for this set of templates {as \textsc{PEGASE} does not include base templates derived using \citet{chabrier2003}.  Nonetheless, we expect differences on stellar masses from using these two IMFs to be small \citep[$M_{\star,\mathrm{Kroupa}} = 1.06 M_{\star,\mathrm{Chabrier}}$, e.g.][]{speagle2014}.} We assume {a uniform dust screen model} using the \citet{calzetti2000} law and with $E(B-V)=0 - 0.3$ with 0.05 mag steps.

\subsection{LePhare}

\lephare\ was originally a photometric redshift code \citep[][]{arnouts1999,ilbert2006}, but it can also be used to estimate a number of physical parameters of galaxies from the best-fit templates. This code is one of the most flexible of those used in this paper, and to minimize differences among different {codes we} use LePhare with the same templates as those described in the previous section (\zpeg\ templates). The only difference is the addition or absence of emission lines on top of the original templates created following the prescription described by \citet{ilbert2006}.

\subsection{MAGPHYS}

\magphys\ \citep{cunha2008} uses stellar templates constructed from the stellar libraries by \citet{bruzual2003} and the dust absorption model {follows \citet{charlot2000}. The adopted IMF is that defined} by \citet{chabrier2003}.  In this code, the star-formation histories are derived from an exponentially declining model and superimposed random bursts. Stellar metallicities are uniformly sampled between 0.02 and 2 times solar metallicity.  {Although there} is no freedom to change the underlying templates,  the code compares the data to the entire library and builds the probability distributions for each physical parameter {(e.g.  stellar mass,  $SFR$, dust, among others)}. Moreover, while this constraint limits our ability to compare directly with other codes, we use a set of libraries that are commonly used in the community and can serve as a standard reference.

\subsection{CIGALE}

\cigale\ is a code that was used to build {optical-to-infrared SED} models with and without AGN contributions that can also be used to estimate physical parameters on {galaxies with no AGN contribution and limited wavelength coverage as is our case} \cite[][]{burgarella2005,noll2009,bouquien2019}. This code allows us a few degrees of freedom, and we try to match the set of available templates to those prescribed by \magphys. The major difference is that we cannot replicate the same star-formation histories, and we use an exponentially delayed $\tau$ model ($SFR(t) \propto t\times \exp\left(-t/\tau\right)/\tau$) with $\tau$ having the same {values between 0.1 and 2 Gyr} as described in Section \ref{ssec:zpeg}.

\subsection{PROSPECTOR}

Finally,  we use \prospector\ \citep[][]{prospector2020,johnson2021} that allows for a Bayesian exploration of the parameter space based on a set of template libraries of choice.  We try our best to mimic the template configuration of \magphys. We allow for the variation of three parameters: stellar-mass (with a top-hat prior $8<\log_{10}(M_\star/\mathrm{M_\odot})<12$); metallicity (with a top-hat prior $-1.7<\log (Z/\mathrm{Z_\odot})<0.3$); and an exponentially declining star formation history with a log-uniform prior $0.1<\tau<30$ Gyr).  The IMF is fixed to that of \citet{chabrier2003}, and we use the dust law defined by \citet{charlot2000} with the dust index fixed at -0.7 (the same as assumed in \textsc{magphys}).


\begin{table*}
    \caption{Median difference [in dex] on estimated stellar masses for all simulated reshifts (one per column) and different codes (one per row) used in this work. Errors are computed as $\sigma/\sqrt{N}$, with $N$ being the number of galaxies in the bin. In the last column we show the overall performance across all redshifts.}              
    \label{table:mass_change}      
    \centering                                      
    \resizebox{\textwidth}{!}{%
        \begin{tabular}{| c | c c c c c c c | c |}          
            \hline
            Code                       & $z=0.1$          & $z=0.3$          & $z=0.5$          & $z=0.8$          & $z=1.0$          & $z=1.5$          & $z=2.0$          & All              \\
            \hline
            \textsc{LePhare}           & -0.03 $\pm$ 0.01 & -0.04 $\pm$ 0.02 & -0.09 $\pm$ 0.02 & -0.15 $\pm$ 0.02 & -0.25 $\pm$ 0.03 & -0.54 $\pm$ 0.03 & -0.70 $\pm$ 0.04 & -0.11 $\pm$ 0.01 \\
            \textsc{LePhare [nolines]} & -0.03 $\pm$ 0.01 & -0.01 $\pm$ 0.02 & -0.02 $\pm$ 0.02 & -0.09 $\pm$ 0.02 & -0.15 $\pm$ 0.03 & -0.35 $\pm$ 0.03 & -0.70 $\pm$ 0.04 & -0.08 $\pm$ 0.01 \\
            \textsc{magphys}           & -0.02 $\pm$ 0.07 & -0.03 $\pm$ 0.06 & -0.08 $\pm$ 0.06 & -0.19 $\pm$ 0.06 & -0.26 $\pm$ 0.06 & -0.31 $\pm$ 0.06 & -0.22 $\pm$ 0.07 & -0.10 $\pm$ 0.03 \\
            \textsc{ZPEG}              & -0.05 $\pm$ 0.02 & -0.06 $\pm$ 0.03 & -0.18 $\pm$ 0.03 & -0.06 $\pm$ 0.03 & 0.01 $\pm$ 0.03  & -0.35 $\pm$ 0.04 & -0.10 $\pm$ 0.06 & -0.12 $\pm$ 0.01 \\
            \textsc{Cigale}            & -0.03 $\pm$ 0.00 & -0.05 $\pm$ 0.01 & -0.16 $\pm$ 0.02 & -0.25 $\pm$ 0.02 & -0.30 $\pm$ 0.02 & -0.39 $\pm$ 0.03 & -0.36 $\pm$ 0.04 & -0.11 $\pm$ 0.01 \\
            \textsc{prospector}        & -0.02 $\pm$ 0.00 & -0.08 $\pm$ 0.01 & -0.14 $\pm$ 0.02 & -0.22 $\pm$ 0.02 & -0.27 $\pm$ 0.02 & -0.23 $\pm$ 0.02 & -0.15 $\pm$ 0.02 & -0.10 $\pm$ 0.01 \\
            \hline
        \end{tabular}}
\end{table*}

\begin{figure*}
    \centering
    \includegraphics[width=\linewidth]{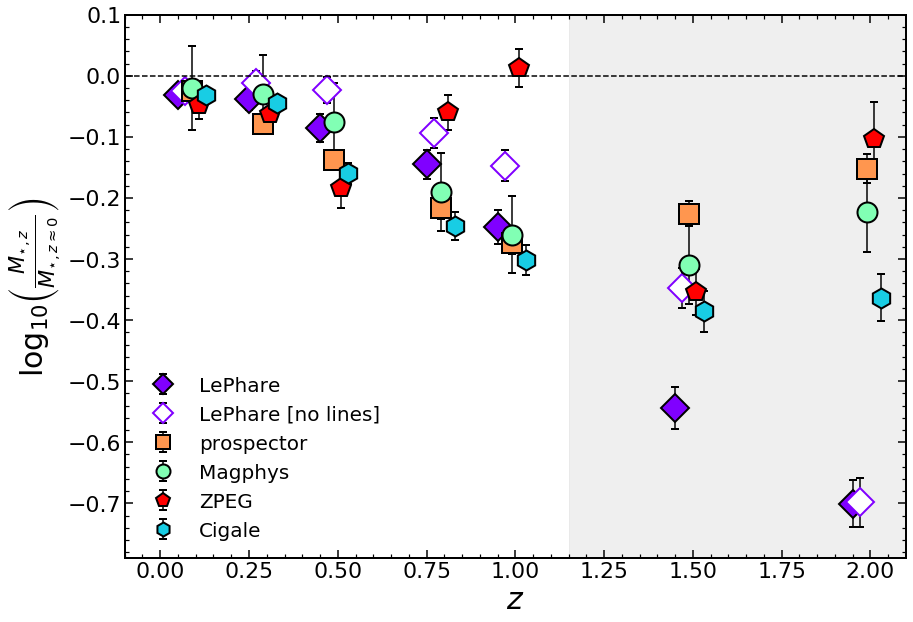}
    \caption{The median stellar mass difference for our sample {of galaxies} as a function of redshift in the case of different SED fitting codes. The stellar mass reference at $z\approx 0$ is computed for each individual code using the same photometric filters. This difference can reach to 0.2-0.3 dex by $z\sim1$.  We find that \textsc{LePhare} (with no added emission lines) gives the best overall results for $z<1$ of all used codes,  but it still underestimates the stellar masses at $z\gtrsim0.5$. The shaded region indicates the redshifts for which the SED fitting codes are not well calibrated, since we are mostly probing regions in the rest-frame UV.}
    \label{fig:dmass_codes}
\end{figure*}

\section{Results and Discussion}\label{sec:results}

{The goal of our work is to study the impact of observational strategies on the derived stellar masses of galaxies.  To test this, we} have applied our artificial redshifting code (\textsc{argas}) to 166 galaxies {from the AMUSING survey} and simulated observations at seven different redshifts $z=0.1, 0.3, 0.5, 0.8, 1.0, 1.5,$ and 2.0. At each redshift, we compute the photometric data in the four $griz$ bands from DECam and use the SED fitting codes described in Section \ref{sec:measure_mass} to get the best stellar mass of the galaxy. {We use as a frame of reference for each code the stellar mass computed at the original redshift of the galaxy ($z\sim0.03$) using the same filters and templates.}

\subsection{Underestimation of stellar masses}\label{ssec:results_masses}
After obtaining our stellar mass estimates, we compare the one obtained at each simulated redshift with the one obtained locally using the same filter set and library templates.  {The median difference for our 166 galaxy sample} between the simulated and local values {are shown} in Fig. \ref{fig:dmass_codes} {(see also Table \ref{table:mass_change})}.

One of the first findings is that, despite the observed differences among the different used codes, there is a systematic underestimation of the stellar mass that depends on the redshift.  This has implications for the implementation of the mass-step correction, as it implies that galaxies which are observed to be below the \msun{10} threshold for correction may actually lie above it.  This effect becomes more prominent as we move towards higher redshifts as more galaxies are affected (larger median offset from the true value). This is an important aspect that needs to be considered when estimating stellar masses for a singular dataset (i.e. observed with the same photometric bands) across an extensive redshift range,  as is the case of large surveys such as DES. {Given our defined set of filters and our choice of stellar population templates,} we find that the \textsc{LePhare} (excluding emission lines) code is the overall best code in estimating stellar masses for galaxies at $z\leq0.5$. Interestingly,  \zpeg\ performs better for galaxies $0.5 < z \leq 1$.  This is likely due to a combination of the nebular emission prescription included in the templates used and the filters where the emission lines are expected to fall. 

Although there are several studies in the literature that tackle a similar issue of estimating physical parameters, they present results using a much broader filter set.  For instance,  \citet{pforr2012},  \citet{mitchell2013}, and \citet{mobasher2015} use optical,  NIR and MIR (IRAC photmetry), \citet{acquaviva2015} use additional UV photometry and more recently \citet{lower2020} uses FIR data from Herschel to constrain physical parameters.  This extended set of photometric points is what is usually required for accurately constraining SED fitting models, given the number of available variables that need constraining  \citep{acquaviva2015, mobasher2015}. Additionally,  none of these studies evaluates the same galaxy simulated at different redshifts. They either consider exclusively mock galaxies \citep[][]{pforr2012, mitchell2013,lower2020}, real data \citep{acquaviva2015} or a mix of both \citep[][]{mobasher2015}. Nevertheless,  the differences among different codes are consistent with results from \citet{mobasher2015}, who found an average spread of 0.136 dex in stellar mass differences estimated from different SED fitting codes using a similar set of assumptions in the model templates. The maximum scatter on the estimation of stellar masses was found to be due to contamination from nebular emission, reaching values of up to 0.5 dex \citep{mobasher2015}. {With respect to stellar mass estimation bias as a function of redshift,} both \citet{pforr2012} and \citet{mitchell2013} find no significant differences.  However, in their test case, they were using a much larger filter set, and estimating stellar masses for mock galaxies simulated to be at the redshift they were being observed.

Interestingly, \citet{pforr2012} tested the impact of assuming different filter sets on photometry of mock galaxies, which includes two sets close to the one we study ($ugriz$ and $UBVRI$).  Contrary to our results, they find no significant difference with redshift, even for these smaller filter sets. We note, however, that galaxies in their study are derived from simulations at the redshift they are being observed, and only include star-forming galaxies with young stars dominating the SED at optical wavelengths. 
{We suppose that it is the fact that we are observing two different types of SEDs at each redshift that is driving the difference between our works. Namely, in our work we use a more evolved population that is the same at all redshifts, whereas in \citet{pforr2012} they simulate star-forming galaxies that evolve with the redshifts they are testing. This tends to counterbalance the effect of filter shifting (when applied over the same population) likely due to a combination of the added $u$-band coverage and a population of younger galaxies. These are two complementary approaches to a similar problem, that nicely test different aspects of stellar mass estimates across large redshift ranges.}

In our experimental set-up,  we are attempting to fit the same galaxies using different rest-frame coverage (here corresponding to the different simulated redshifts) and a small filter set for SED fitting {to mimic the conditions for large sky surveys where most SN are found}.  We find that the one feature that most affects the measured stellar mass is the possibility to constrain the 4000\AA\ break that allows one to have an idea of the fraction of young and old stars in the galaxy and better constrains the average stellar population age.  As we move towards higher redshifts, we are sampling increasingly bluer wavelengths, and thus {giving more weight to the younger stellar population \citep[e.g.][]{pforr2012, mobasher2015} that can outshine the older stellar populations which add up to most of the galaxy stellar mass \citep[especially in star-forming galaxies, see e.g.][]{sorba2018}}.  And since these stars have lower mass-to-light ratios,   estimates of stellar masses based on these wavelengths tend to be lower than the true value \citep[][]{pforr2012}.

\begin{figure}
    \centering
    \includegraphics[width=\linewidth]{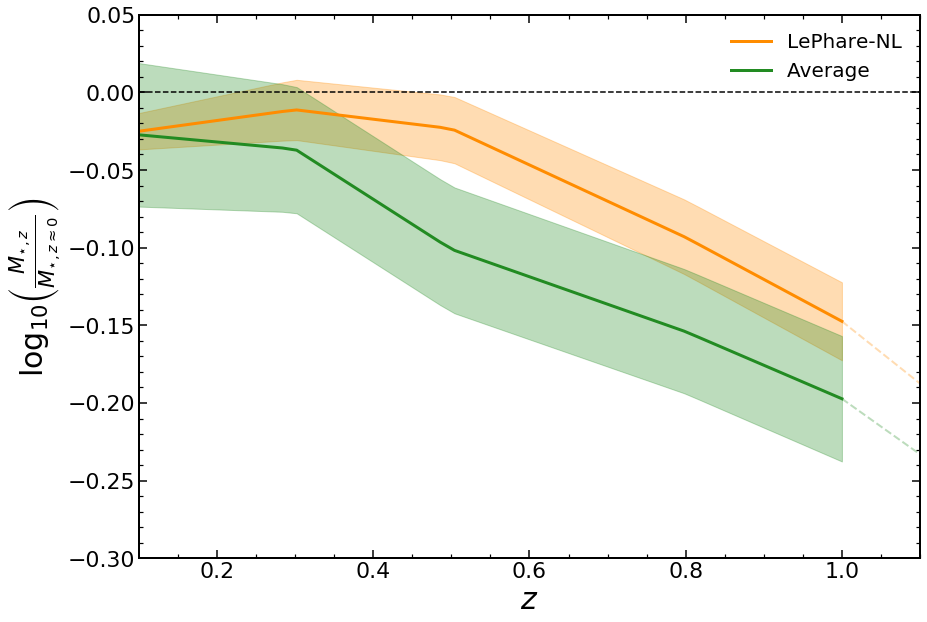}
    \caption{The mass correction function for each redshift to be applied to the {estimated} observed stellar masses.  The orange line represents the global mass correction using our best approximation (\lephare\ [no-lines]).  The green line shows the correction to be applied using the weighted average correction derived from all SED fitting codes. The shaded regions represent the uncertainty on the correction at each redshift.}
    \label{fig:corrections}
\end{figure}


\begin{figure}
    \centering
    \includegraphics[width=\linewidth]{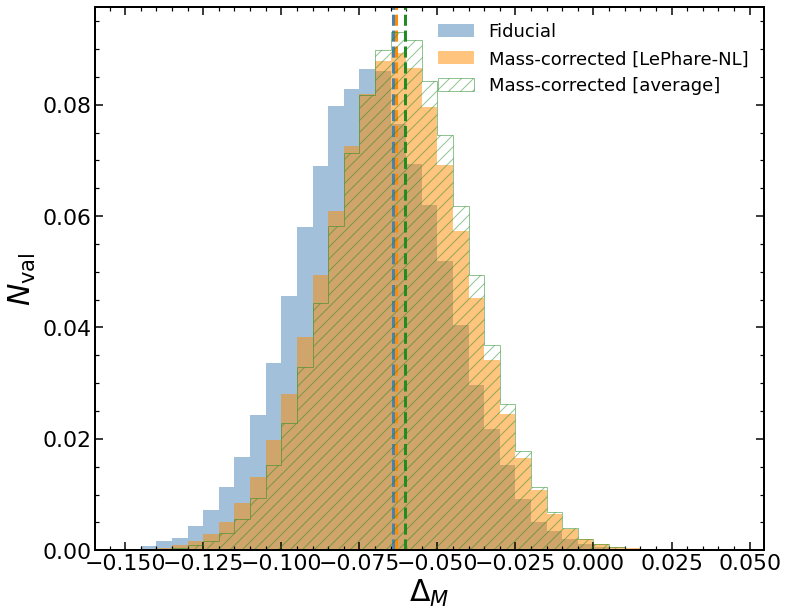}
    \caption{The resulting posterior distributions on $\Delta_M$ for different runs.  We show the \textsc{LePhare-NL} stellar mass correction results in orange and average stellar mass correction results in green,  compared to the fit using the original stellar masses from \citet{betoule2014}.  Vertical lines indicate the best-fit value for each configuration. {We find that the fiducial model has a slightly larger value for $\Delta_M$ than either of the mass-corrected models, being very close to the best-fit value for the \textsc{LePhare-NL} correction model. Nevertheless, the resulting distributions for both mass-corrected models are similar among themselves.}}
    \label{fig:DeltaM_dist}
\end{figure}

\begin{figure}
    \centering
    \includegraphics[width=\linewidth]{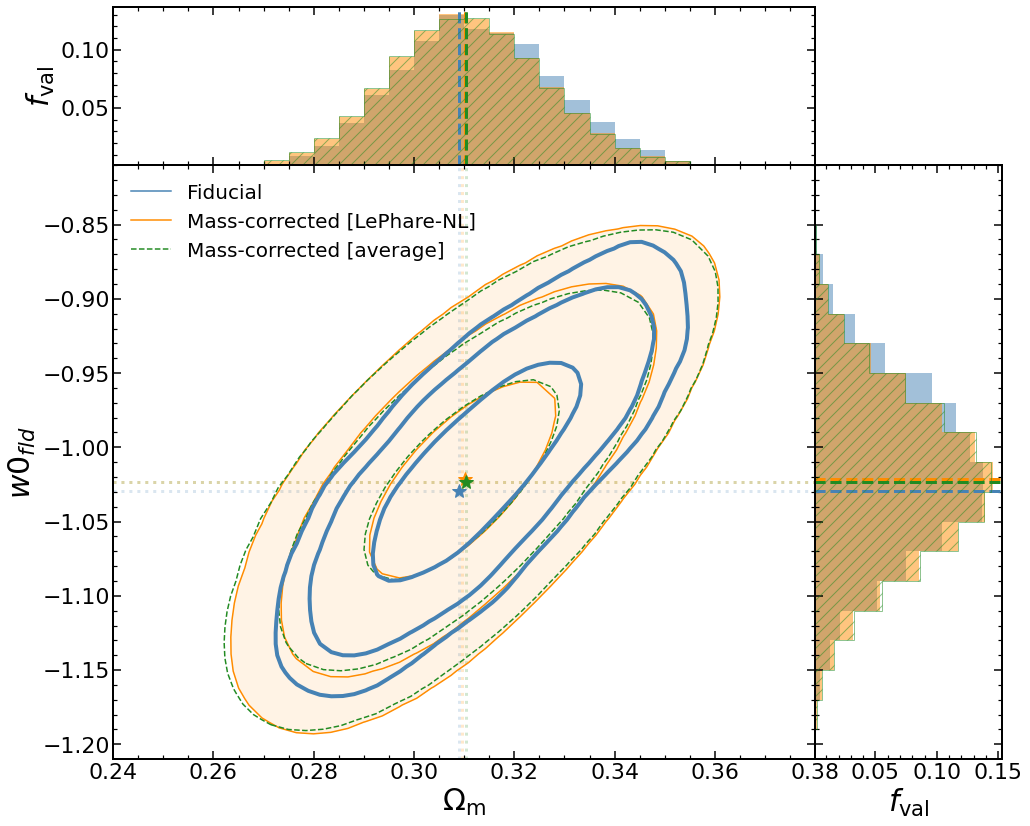}
    \caption{The resulting posterior distributions on $\omega$ and $\Omega_m$ for different stellar mass corrections (\textsc{LePhare-NL} stellar mass correction in orange and average stellar mass correction in green), compared to the fit using the original stellar masses from \citet{betoule2014}, in blue.  The contours levels correspond,  from inside out, to 68\%, 95\% and 99\% of the posterior distribution. We show as stars (same colours as contours) the best-fit value for each configuration.  Vertical and horizontal lines indicate the best-fit value of each parameter for the three different set-ups. There is no significant difference on the constraints when using the mass-corrected dataset with respect to the fiducial model.  We find small differences in the best-fit values (star symbols), with the \textsc{LePhare-NL} stellar mass correction configuration showing the largest difference with respect to the fiducial model.}
    \label{fig:w_omega_dist}
\end{figure}

\begin{table*}
    \caption{Best-fit parameters of the cosmological model based on the three different configurations described in Sect. \ref{ssec:results_cosmo}. }              
    \label{table:cosmo_results}      
    \centering                                      
    \begin{tabular}{c c c c}          
        \hline
        Parameter     & Fiducial                    & Mass-corrected [LePhare-NL]    & Mass-corrected [average]    \\
        \hline
        $w0_{fld }$ & -1.029$^{+0.069}_{-0.043}$ & -1.022$^{+0.053}_{-0.056}$ & -1.023$^{+0.053}_{-0.055}$\\
        $\alpha$ & 0.142$^{+0.006}_{-0.007}$ & 0.141$^{+0.007}_{-0.006}$ & 0.141$^{+0.007}_{-0.006}$\\
        $\beta$ & 3.080$^{+0.076}_{-0.083}$ & 3.070$^{+0.087}_{-0.074}$ & 3.068$^{+0.087}_{-0.075}$\\
        $M$ & -19.108$^{+0.036}_{-0.043}$ & -19.110$^{+0.040}_{-0.037}$ & -19.111$^{+0.041}_{-0.037}$\\
        $\Delta_{M }$ & -0.064$^{+0.016}_{-0.030}$ & -0.063$^{+0.022}_{-0.023}$ & -0.060$^{+0.020}_{-0.022}$\\
        $\Omega_{m }$ & 0.309$^{+0.019}_{-0.012}$ & 0.310$^{+0.016}_{-0.015}$ & 0.310$^{+0.015}_{-0.015}$\\
        $H0$ & 68.249$^{+1.248}_{-2.034}$ & 68.075$^{+1.594}_{-1.600}$ & 68.043$^{+1.652}_{-1.546}$\\
        \hline                                             
    \end{tabular}
\end{table*}

\subsection{Impact on cosmology}\label{ssec:results_cosmo}

We find that {galaxy} stellar mass corrections {depend strongly} on the observed redshift. This can be a problem for cosmological fits based on SN data that spans a sizeable cosmic time and use {SN host galaxies stellar masses} as the third empirical correction to their brightness. Our findings imply that some galaxies observed at stellar masses lower than $10^{10}\mathrm{M_\odot}$ are more likely to {actually be above that correction threshold.  This} is increasingly critical as we move towards higher redshifts.  In this sub-section, we use our derived corrections to estimate their impact on the derived best-fit cosmological models.

We use the median stellar mass difference to re-estimate the best-fit cosmological parameters for the \citet{betoule2014} sample. We do this using two different approaches. The first using the best approximation we derive from the set of SED fitting codes that were tested in our paper (i.e. \lephare\ [no-lines]).  In the second approach we combine all individual corrections using a weighted average to produce a global correction curve for the estimated stellar masses. To derive the stellar mass correction curve as a function of redshift ($\Delta M(z)$), we interpolate linearly between the simulated redshifts. We show these correction curves in Fig. \ref{fig:corrections}. {We restrict our stellar mass corrections to be valid only at $z\leq1$. This has negligible impact on our tests for cosmological parameters since the majority of SNe are below that redshift limit.}

The distance modulus to each supernova can be modelled as \citep[e.g.][]{betoule2014}:
\begin{equation}
    \mu(z) = M_B + 5\log_{10}(H_0[z, \Omega_m, w]) - \alpha \times (s - \bar{s}) + \beta \times c
\end{equation}
where $s$ is the stretch term and $c$ is the colour term.  The supernova luminosity is parameterized by
\begin{equation}
    M_B =
    \begin{cases}
        M_{B,1} + \Delta_M, & \text{if } M_\star\geq 10^{10}\mathrm{M_\odot} \\
        M_{B, 1},           & \text{otherwise}
    \end{cases}
    \label{eq:mass_step}
\end{equation}
with $M_{B, 1}$ being a free parameter, and $\Delta_M$ the magnitude difference to be applied for SNe Ia in massive hosts.

We estimate the best-fit parameters and corresponding probability density distributions using an MCMC approach with the package \textsc{MontePython} \citep[][]{Brinckmann2018, Audren2012}.  Our analysis is conducting using the "Joint Light-Curve Analysis" sample \citep[][\!, hereafter referred to as JLA]{betoule2014}. This sample combines data from 740 SNe Ia up to redshift $z\sim1.3$ and data from the cosmic microwave background \citep[CMB][]{planck2018}. We use a likelihood defined as \citep[see e.g.][]{gaitan2020}:
\begin{equation}
    -2\,\mathrm{ln}(\mathcal{L})=\sum_\mathrm{SN} \left \{ \frac{\left [\mu(z) - \mu_\mathrm{obs} \right ]^{2}}{\sigma^{2}_\mathrm{tot}} \right \} ,
    \label{eq:likelihood}
\end{equation}
where the uncertainty is defined as the diagonal of the covariance matrix:
\begin{equation}
    \sigma^{2}_\mathrm{tot}=\sigma^{2}_{m_{B}} + (\alpha \sigma_{S})^2 + (\beta \sigma_{C})^2+\sigma^{2}_\mathrm{int}.
\end{equation}
We assume that $\sigma_\mathrm{int}=0.105$, which is the average value for the JLA sample. We use the constraints of CMB data as a prior in our model in the same functional form as in equation 18 by \citet{betoule2014}, only updating the values with the latest release from the Planck survey \citep[][]{planck2018}.

{To incorporate the uncertainty of the stellar mass correction models (see Fig. \ref{fig:corrections}), we create 50 different corrections curves that are randomly perturbed around the median correction, and within the shown uncertainty region. We then run our cosmological fits for each of the 50 individual corrections. Finally, we combine the results into a single posterior distribution for each parameter, marginalized over the uncertainty on the stellar mass correction.}

We do this exercise in three different configurations: one using the original stellar masses from the JLA sample, which is our fiducial model; and the two other MCMC runs are using the derived stellar mass corrections shown in Fig. \ref{fig:corrections} applied to the measured stellar masses of the JLA sample.   The best-fit values and corresponding uncertainties for each of these configurations is shown in Table \ref{table:cosmo_results}.  We also show all the posterior distributions for the fitted parameters in Fig.  \ref{fig:corner_plot}.

We find that the parameters that changes the most when applying our stellar mass corrections is $\Delta_M$.  In our \lephare -corrected model, the best-fit parameter value decreases by $\sim$2\%,  while when we apply our average-correction the difference with respect to the fiducial model is $\sim$6\%.  Since $\Delta_M$ is the parameter that is linked on the host stellar mass (Eq. \ref{eq:mass_step}), it is expected that it is the most affected by applying corrections to the original stellar masses (see Fig \ref{fig:DeltaM_dist}).

With respect to the constraints of the cosmological parameters, we find smaller differences with respect to the fiducial model ($<1$\%, see Figures \ref{fig:DeltaM_dist} and \ref{fig:w_omega_dist}).  The value of $\Omega_m$ increases by $\delta \sim 0.001$ ($\sim0.3$\%) in the \lephare -corrected and average-corrected fits.  The value of $w_0$ increases by $\delta \sim 0.007$ ($\sim0.7$\%) when using the \lephare\ stellar mass corrections. To contextualize, this possible systematic bias corresponds to approximately one tenth of the expected error budget in $w_0$ from Euclid \citep[][]{amendola2018}. The difference in $w$ is slightly smaller when we use the average correction, with it increasing only by $\delta \sim 0.006$ ($\sim0.6$\%).  Finally, we find also small changes in the $H_0$ parameter: $\delta\sim-0.2$ ($\sim-0.3$\%) for both the \lephare\ and the average stellar mass corrections.  We note, however, that these differences are all within the fitting uncertainties.


\section{Conclusions}\label{sec:conclusions}

We study the impact of observational effects (namely cosmological dimming and rest-frame coverage) on estimating physical parameters  of galaxies. In particular, we aimed to assess possible systematic bias on the estimation of stellar masses when analysing galaxy samples across a large redshift range. 
To achieve this goal we use a sample of 166 SNe Ia host galaxies with IFS from the AMUSING survey. With these galaxies it was possible to simulate observations of galaxies at redshifts ($0.1<z<2$ using a $griz$ filter set to mimic the DES-SN program). Five different codes --- \cigale, \lephare, \magphys, \prospector, \zpeg\ --- were used to estimate stellar masses allowing for a better identification of possible bias associated with the choice of SED fitting models.  The implications of our results on the determination of cosmological parameters using mass step correction has been studied. Our main conclusions are:

\begin{itemize}
    \setlength\itemsep{0.5em}
    \item Regardless of the code used to estimate stellar masses,  there is a systematic underestimation of the stellar mass,  which increases with increasing redshift.  Depending on the individual code, this difference reaches around 0.2-0.3 dex by $z\sim1$.
    \item We find that when correcting the observed stellar masses for a public SNe Ia sample, there is a small impact on the best-fit parameters of the cosmological model. The impact has the same order of magnitude whether we use the \textsc{LePhare-NL} or the average stellar mass corrections.
    \item {The cosmological parameters show the largest impact when deriving the best-fit value of the magnitude correction $\Delta_M$, which is reduced by $\sim$2 and $\sim$6\% for the \textsc{LePhare-NL} and average stellar mass corrections, respectively.  The cosmological parameters show deviations from the fiducial value below 1\%: $\Omega_m$ increases by \deltaOmpercent\% ($\delta \sim $ \deltaOm); $w$ is reduced by \deltawpercent{}\% ($\delta \sim $ \deltaw); and $H_0$ decreases by \deltaHpercent\% ($\delta \sim $ \deltaH). These differences are all within the fitting uncertainties,  but could be a non negligible source of systematic errors in the coming decade.}
\end{itemize}

Our main conclusion is that stellar mass estimations across a large redshift range have a systematic underestimation that itself depends on the redshift of the observed host galaxy. The forthcoming surveys, such as Euclid and/or Nancy Grace Roman Space Telescope, can help minimize these effects by providing a more significant baseline of rest-frame coverage (with added filters in the NIR regime) that helps minimize the error budget.  By doubling the number of filters into the NIR regime,  one can hope to constrain better the region around 4000\AA\ rest-frame to higher redshifts, helping quantify the amount of old and young stars in the galaxy, which are crucial for accurate stellar mass estimates.


\begin{acknowledgements}
    This work was supported by Funda\c{c}\~{a}o para a Ci\^{e}ncia e a Tecnologia (FCT) through the research grant UIDB/00099/2020 and the CRISP project PTDC/FIS-AST-31546/2017.
    L.G. acknowledges financial support from the Spanish Ministry of Science, Innovation and Universities (MICIU) under the 2019 Ram\'on y Cajal program RYC2019-027683 and from the Spanish MICIU project HOSTFLOWS PID2020-115253GA-I00.
    CRA was supported by grants from VILLUM FONDEN (project numbers 16599 and 25501).
    JDL acknowledges support from a UK Research and Innovation Future Leaders Fellowship (MR/T020784/1).

    Computations were performed at the cluster "Baltasar-Sete-Sóis", supported by the H2020 ERC Grant "Matter and strong field gravity: New frontiers in Einstein's theory" grant agreement no. MaGRaTh-646597, and at COIN, the CosmoStatistics Initiative, whose purchase was made possible due to a CNRS MOMENTUM 2018-2020 under the project "Active Learning for large scale sky surveys”.

    This work was only possible by the use of the following \textsc{python} packages: NumPy \& SciPy \citep{walt2011,jones2001}, Matplotlib \citep{hunter2007}, and Astropy \citep{robitaille2013}.
\end{acknowledgements}


\appendix

\section{Full results from cosmological fits}

In this section we show the posterior distributions for all the fitted parameters in our \textsc{MontePython} model (see description in Sect. \ref{ssec:results_cosmo}). In Fig. \ref{fig:corner_plot} we show that for most parameters the distributions are similar, with the variable that shows the largest impact from correcting the stellar masses being $\Delta_M$, the magnitude correction to be applied depending on the host stellar mass (see Eq. \ref{eq:mass_step}).

\begin{figure*}
    \centering
    \includegraphics[width=\linewidth]{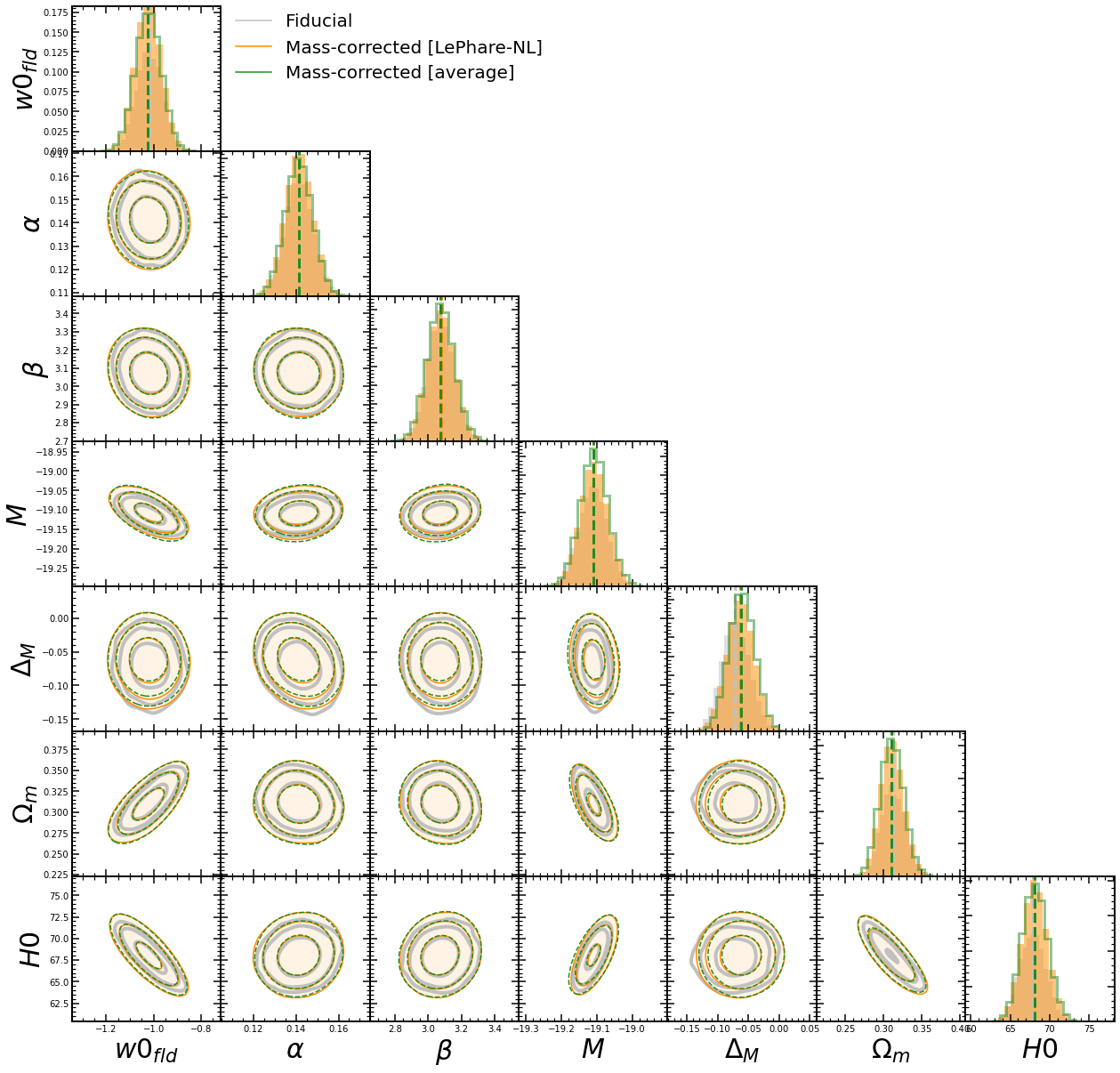}
    \caption{The resulting posterior distributions on all the free parameters for our cosmological model using the three different configurations: fiducial model (in gray),  stellar masses corrected using best approximation with \textsc{LePhare-NL} (in orange), and stellar masses corrected with the average difference among different codes (in green). The contours levels correspond,  from inside out, to 68\%, 95\% and 99\% of the posterior distribution. }
    \label{fig:corner_plot}
\end{figure*}

\bibliographystyle{aa}
\bibliography{refs}

\end{document}